\documentclass[aps,prl,twocolumn,superscriptaddress,nofootinbib,nobibnotes,longbibliography]{revtex4-2}

\usepackage{amsmath,amssymb}
\usepackage{graphicx}
\usepackage{hyperref}
\graphicspath{{figure/}{figures_ED/}{figures/}{figures_extended/}{./}}
\usepackage{xcolor}
\hypersetup{
    colorlinks=true,
    linkcolor=blue,
    citecolor=blue,
    urlcolor=blue
}
\usepackage{bm}\usepackage{mathtools}
\newcommand{\FloatBarrier}{\par}

\newcommand{\mydel}[1]{}

\newcommand{\methodname}{AMVP}
\newcounter{amvpalgorithm}
\renewcommand{\theamvpalgorithm}{\arabic{amvpalgorithm}}
\renewcommand{\equationautorefname}

\begin{document}

\title{Gradient-estimator design overcomes trainability barriers in neural-network-based variational optimization}

\author{Yi-Ran Xue}
\affiliation{National Laboratory of Solid State Microstructures and Department of Physics, Nanjing University, Nanjing 210093, China}
\affiliation{Department of Physics, University of Massachusetts, Amherst, Massachusetts 01003, USA}

\author{Rui Wang}
\email[Corresponding author: ]{rwang89@nju.edu.cn}
\affiliation{National Laboratory of Solid State Microstructures and Department of Physics, Nanjing University, Nanjing 210093, China}
\affiliation{Collaborative Innovation Center of Advanced Microstructures, Nanjing University, Nanjing 210093, China}
\affiliation{Jiangsu Physical Science Research Center, Nanjing 210093, China}
\affiliation{Hefei National Laboratory, Hefei 230088, People's Republic of China}

\author{Baigeng Wang}
\email[Corresponding author: ]{bgwang@nju.edu.cn}
\affiliation{National Laboratory of Solid State Microstructures and Department of Physics, Nanjing University, Nanjing 210093, China}
\affiliation{Collaborative Innovation Center of Advanced Microstructures, Nanjing University, Nanjing 210093, China}
\affiliation{Jiangsu Physical Science Research Center, Nanjing 210093, China}

\author{Chenan Wei}
\email[Corresponding author: ]{chenanwei@umass.edu}
\affiliation{National Laboratory of Solid State Microstructures and Department of Physics, Nanjing University, Nanjing 210093, China}
\affiliation{Department of Physics, University of Massachusetts, Amherst, Massachusetts 01003, USA}
\affiliation{A. Alikhanyan National Science Laboratory, Br. Alikhanian 2, Yerevan 0036, Armenia}
\affiliation{Institute of Theoretical Physics, Faculty of Physics, University of Warsaw, Warsaw, Poland}

\date{\today}

\begin{abstract}
Neural networks provide expressive representations for scientific computing.
However, even sufficiently expressive networks can suffer training failure in weak-gradient regimes, limiting their practical use in quantum many-body physics and \textit{ab initio} quantum chemistry.
Here we derive an unbiased direct gradient estimator and introduce the adaptive minimum-variance phase (AMVP) estimator for neural-network variational optimization. 
By improving the signal-to-noise ratio of weak gradients, these methods enable reliable scientific calculations where training previously failed, while substantially reducing computational cost.
The framework enables compact networks to outperform larger and fine-tuned default standard-estimator models with over an order of magnitude less GPU time on correlated flux models, and ultimately exceed the density matrix renormalization group (DMRG) accuracy.
It further achieves chemical accuracy in N$_2$ bond breaking and, for the first time, in heavy-element I$_2$ with explicit spin–orbit coupling.
These results demonstrate that gradient-estimator design expands the capabilities of neural-network variational methods for accurate scientific computing.

\end{abstract}

\maketitle


\textbf{Introduction.}\enspace Neural networks provide expressive representations for scientific computing\cite{gao2017efficient}.
Across quantum many-body physics and \textit{ab initio} quantum chemistry, neural-network variational methods have shown great potential in representing complex wavefunctions in challenging systems\cite{carleo2017solving,pfau2020abinitio,hermann2020deep,medvidovic2024neural,tang2025deep}.
However, greater network expressivity does not by itself ensure successful optimization: training can still fail despite extensive fine-tuning and substantial computational investment, limiting the practical application of these methods\cite{bukov2021learning,hermann2023abinitio,ledinauskas2025universal}.

Advances in sampling and optimization have helped neural-network variational Monte Carlo (NN-VMC) better exploit the expressive power of wavefunctions. For example, autoregressive models ease sampling bottlenecks through independent Born sampling\cite{sharir2020deep}, while minimum-step stochastic reconfiguration (MinSR) enables accurate training of deep, expressive networks\cite{chen2024empowering}. 
Despite these advances, training can still fail in complex systems when weak descent signals are overwhelmed by gradient noise\cite{sinibaldi2023unbiasing,misery2026looking}.
In particular, one of the challenging cases is learning the signs and phases that govern quantum interference in strongly correlated and spin–orbit-coupled (SOC) systems\cite{westerhout2020generalization,szabo2020neural,bukov2021learning,melton2016variable}.

In this work, we tackle such training failures in NN-VMC through estimator design. This principle has shown its ability to recover effective optimization signals in machine learning\cite{ranganath2014black,parmas2018pipps,grathwohl2018backprop,tucker2019doubly}, suggesting an overlooked route to successful NN-VMC calculations. Specifically, we derive an unbiased direct gradient estimator and further introduce the adaptive minimum-variance phase (AMVP) estimator, which improves the signal-to-noise ratio of weak phase gradients. This framework broadens the practical scope of NN-VMC, enabling reliable and accurate calculations for systems previously inaccessible at significantly lower computational cost.

\begin{figure*}[t]
\centering
\includegraphics[width=\textwidth]{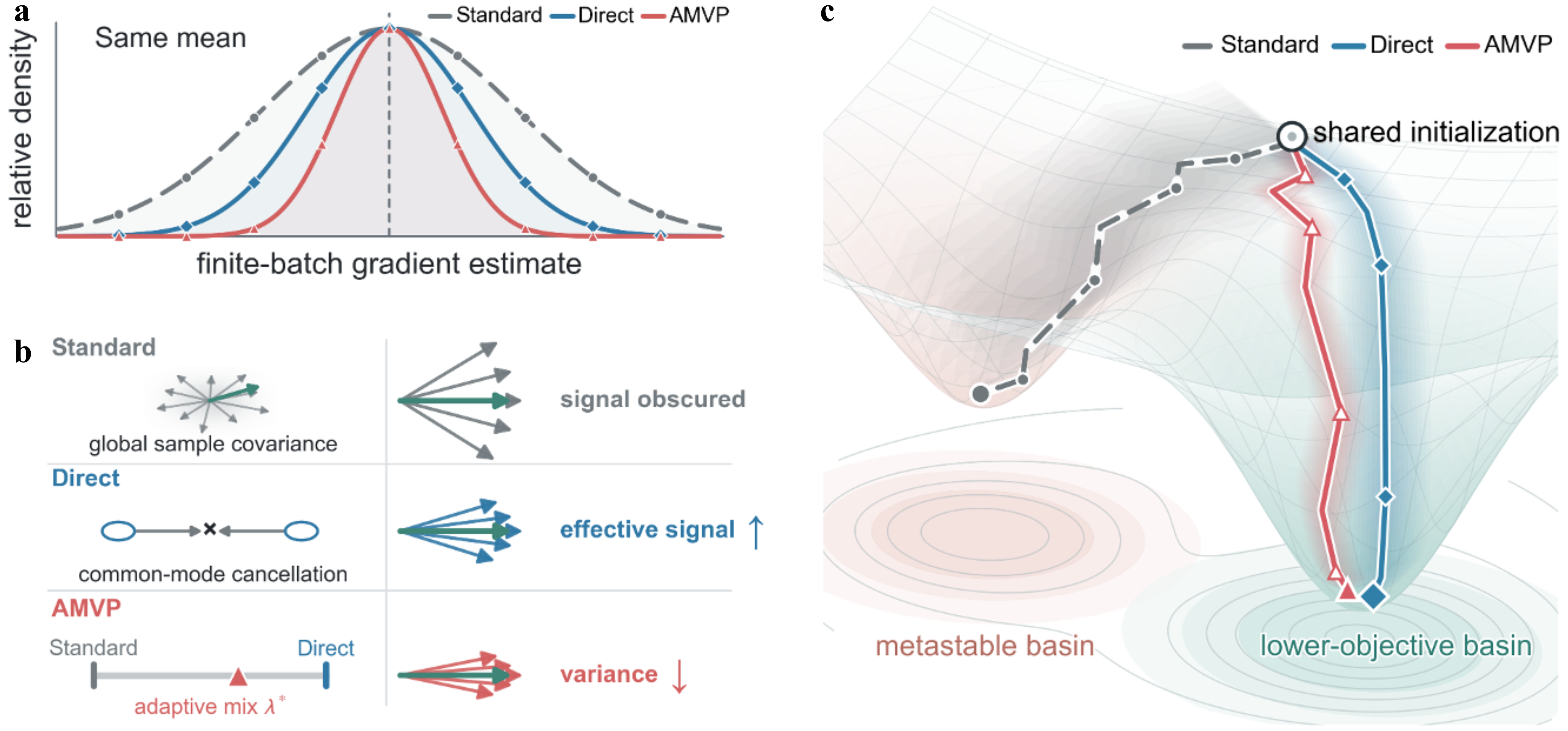}
\caption{
\textbf{Estimator design changes the stochastic optimization dynamics.}
(a) The standard, direct, and adaptive minimum-variance phase
(AMVP) estimators share the same population gradient but exhibit different finite-sample fluctuations. 
(b) The standard estimator first aggregates the imaginary local-energy contributions before coupling them to the phase score, whereas the direct estimator retains connection-resolved phase differences that remove common-mode fluctuations.
AMVP adaptively combines unbiased estimators according to their stochastic properties.
(c) Different estimator noise structures lead to different optimization trajectories. The standard estimator can leave optimization trapped in
metastable regimes, whereas the direct and AMVP estimators allow neural states to better exploit their available capacity and approach target states.
}
\label{fig:estimator_dynamics}
\end{figure*}

We first demonstrate these capabilities in synthetic-flux quantum systems motivated by ultracold-atom realizations\cite{atala2014observation,liang2024chiral,impertro2025strongly}. On the ladder, a compact network reaches subpercent energy error, far outperforming standard-estimator ResNet and multilayer perceptron (MLP) baselines across learning-rate, width and depth sweeps, with an order of magnitude less GPU time. On the two-dimensional torus, AMVP outperforms our density matrix renormalization group (DMRG)\cite{white1992density} benchmark in median variational energy. Furthermore, we overcome intrinsic sign and phase barriers from N$_2$ bond breaking to SOC\cite{melton2016variable}. For the heavy-element molecule I$_2$, we achieve, to our knowledge, the first NN-VMC optimization with explicit SOC: most runs succeed in reaching chemical accuracy, while all standard runs fail.

To explain these gains, we combine theoretical analysis with matched-state numerical diagnostics, linking reduced gradient variance to successful sign and phase learning in correlated matter and relativistic quantum chemistry. More broadly, these findings establish gradient-estimator design as a core component of scientific computation with neural networks.

\textbf{Results.}\enspace 

\textbf{Gradient estimators define distinct stochastic optimization dynamics.}\enspace At finite sample sizes, statistical fluctuations can obscure weak descent signals and impede optimization. Training dynamics depend on both the population gradient and the finite-sample fluctuations of its estimator. For an unbiased estimator,
\begin{equation}
\hat{g}=\nabla E+\xi ,
\label{eq:grad_decomposition}
\end{equation}
where $\nabla E$ denotes the population gradient and $\xi$ represents finite-sample fluctuations. Different estimators can therefore target the same
variational gradient while generating different optimization trajectories\cite{mohamed2020monte}.

This distinction becomes particularly important in neural-network variational methods. Specifically, in neural-network quantum states (NQS), this challenge is especially pronounced in the phase sector, where effective optimization signals can be weak, while estimator fluctuations remain substantial. Here, we compare the standard score-function estimator and a direct phase-gradient estimator while keeping the ansatz, sampling procedure, and optimization protocol unchanged.
Although both estimators are unbiased for the same variational force, $\left\langle \hat{g}_{\mathrm{std}}\right\rangle
=
\left\langle \hat{g}_{\mathrm{dir}}\right\rangle
=
\nabla E ,$ they generate different finite-sample stochastic dynamics.

For $\Psi_{\theta}(R)=\exp[u_{\theta}(R)+\mathrm{i}\varphi_{\theta}(R)]$, a phase-only parameter $\beta$ satisfies $\partial_{\beta}u_{\theta}=0$, and its conventional score estimator is
$\hat{g}^{\mathrm{std}}_{\beta}(R)
=2\,\operatorname{Im}E_{\mathrm{loc}}(R)\,
\partial_{\beta}\varphi_{\theta}(R)$. When amplitude and phase share parameters, the bare local-energy derivative omits the amplitude contribution. We therefore retain the standard amplitude force and replace only the phase channel by its direct form,
\begin{equation}
\begin{aligned}
\hat{\boldsymbol{g}}^{\mathrm{dir,cpl}}_{\theta}(R)
={}&
2\left[
\operatorname{Re}E_{\mathrm{loc}}(R)-E
\right]
\nabla_{\theta}u_{\theta}(R)
\\
&+
\sum_{R'}
\operatorname{Im}\!\left[W_{RR'}\right]
\left[
\nabla_{\theta}\varphi_{\theta}(R)
-
\nabla_{\theta}\varphi_{\theta}(R')
\right],
\end{aligned}
\label{eq:coupled_direct_estimator}
\end{equation}
where $W_{RR'}=H_{RR'}\Psi_{\theta}(R')/\Psi_{\theta}(R)$, and gradients through $u_{\theta}$ are stopped in the direct phase term. This construction preserves the full population gradient while allowing connection-resolved phase-score fluctuations to cancel before summation.

The AMVP dynamically combines the two unbiased phase estimators,
\begin{equation}
\hat{g}_{\lambda}
=
(1-\lambda)\hat{g}_{\mathrm{std}}
+
\lambda\hat{g}_{\mathrm{dir}},
\label{eq:amvp_combination}
\end{equation}
where the mixing coefficient is selected to minimize the measured gradient variance\cite{greensmith2004variance,metz2019understanding} (see Methods and Supplementary Information Sec.~4),
\begin{equation}
\lambda^\star
=
\arg\min_{\lambda\in[0,1]}
\mathrm{Var}(\hat{g}_{\lambda}).
\label{eq:amvp_optimization}
\end{equation}

As illustrated in Fig.~\ref{fig:estimator_dynamics}, the AMVP estimator exploits the freedom of unbiased estimator design\cite{assaraf1999zero} to select the stochastic channel with lower gradient fluctuations. Here, Hermiticity yields a direct local-energy derivative estimator of the phase force that shares the population mean of the standard score estimator but has different finite-sample fluctuations. In coupled amplitude--phase networks, parameter sharing further introduces amplitude--phase cross-covariances, so AMVP minimizes the variance of the full stochastic force rather than that of an isolated phase estimator.

Furthermore, by suppressing estimator-induced stochastic fluctuations, AMVP stabilizes optimization, allowing neural states to realize more of their available expressive capacity and more accurately approximate optimal states without increasing architecture complexity or extensive hyperparameter tuning.

\textbf{Overcoming phase-optimization barriers in quantum flux models.}\enspace To examine whether NQS with complex phase structures can fully exploit their available variational capacity, we first consider a flux ladder model. The Peierls phases generate chiral currents and introduce non-trivial phase structures into the wavefunction\cite{orignac2001meissner,hugel2014chiral,atala2014observation}, making the system sensitive to the quality of phase-gradient estimation. For the flux ladder benchmark, we use separated amplitude--phase networks, which allow the phase-gradient estimator to be isolated directly. This setting provides a controlled test of whether the estimator choice alone can alter the optimization outcome. The Hamiltonian is given by
\begin{align}
H_{\mathrm{ladder}}
=&
\sum_{\langle ij\rangle\in \mathrm{legs}}
\left[
J_z S_i^z S_j^z
+
\frac{J_{\mathrm{leg}}}{2}
\left(
e^{\mathrm{i}A_{ij}}S_i^+S_j^-+\mathrm{h.c.}
\right)
\right]
\nonumber\\
&
+\sum_{\langle ij\rangle\in \mathrm{rungs}}
\left[
J_z S_i^z S_j^z
+
\frac{J_{\mathrm{rung}}}{2}
\left(
S_i^+S_j^-+\mathrm{h.c.}
\right)
\right],
\label{eq:flux_ladder_ham}
\end{align}
where $A_{ij}=+\Phi/2$ on one leg and $A_{ij}=-\Phi/2$ on the other leg in the symmetric gauge. We use
$J_{\mathrm{leg}}=1$, $J_{\mathrm{rung}}=0.8$, and $J_z=0.5$.

\begin{figure*}[t]
\centering
\includegraphics[width=\textwidth]{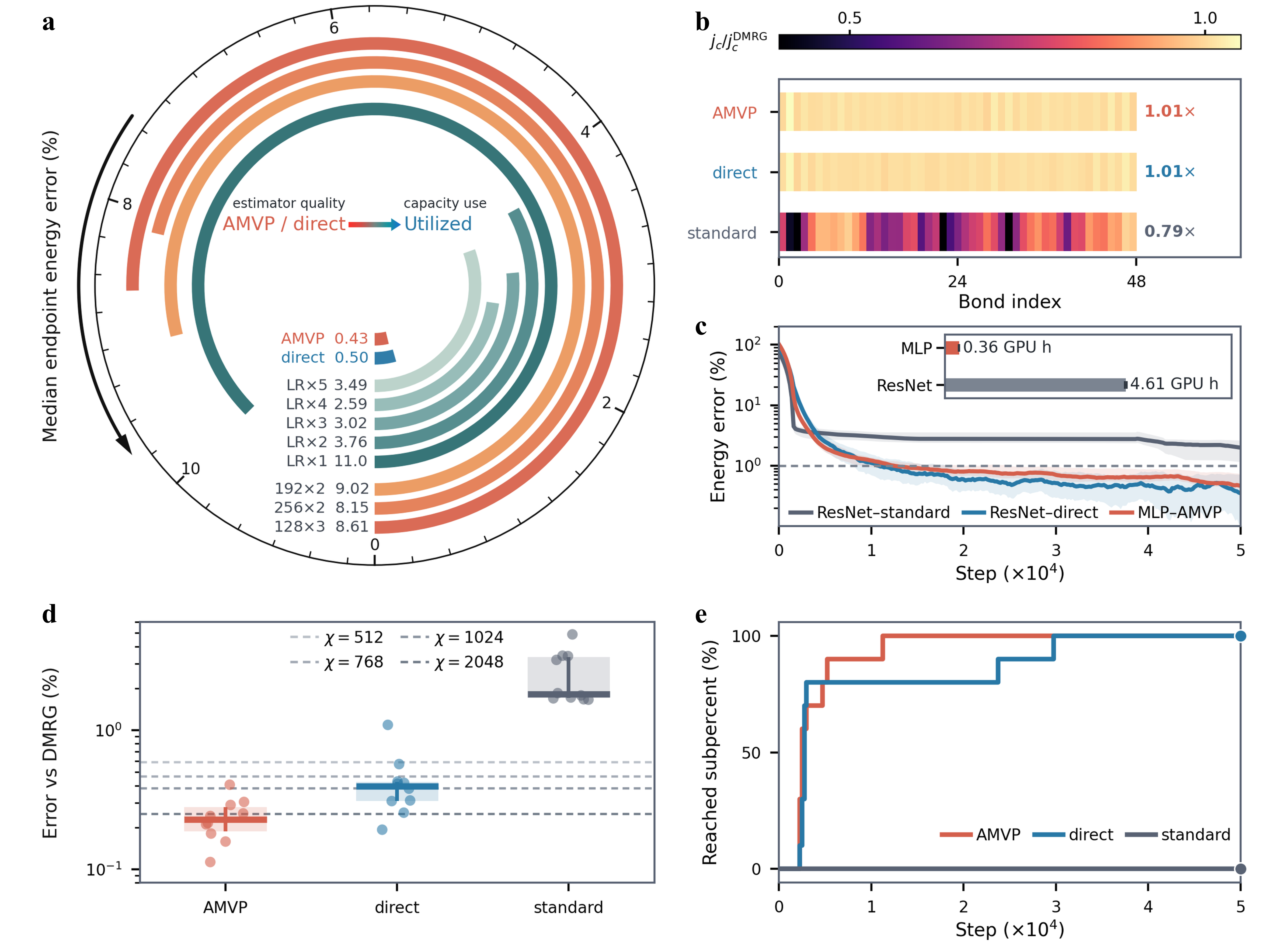}
\caption{
\textbf{Gradient-estimator choice overcomes a phase-optimization barrier.}
(a) On a 50-rung flux ladder, the standard estimator remains trapped in a high-error regime despite
variations in learning rate and network capacity. In contrast, the direct and adaptive minimum-variance phase (AMVP) estimators allow the same neural-network representation to reach subpercent errors, with median errors of $0.50\%$ and $0.43\%$, respectively. (b) The chiral current profile obtained with direct and AMVP agrees with the DMRG reference ($j_c/j_c^{\mathrm{DMRG}}\approx1.01$), whereas the standard estimator shows a marked deviation ($j_c/j_c^{\mathrm{DMRG}}\approx0.79$).
(c) Training trajectories and computational cost. Both a direct-trained ResNet and a compact AMVP-trained MLP reach energy errors below 0.5\%, whereas the standard-trained ResNet remains in a higher-error regime. The inset compares allocated A100 GPU times for AMVP-trained MLP and standard-trained ResNet runs: 0.36 GPU h and 4.61 GPU h, respectively.
(d) Energy errors for an $8\times8$ flux square with periodic boundary conditions. Energies obtained with AMVP estimators are lower than the finite-$\chi$ DMRG energies obtained up to $\chi=2048$, whereas the standard estimator remains in a higher-error regime.
(e) Cumulative fraction of runs that have reached subpercent energy error.
}
\label{fig:flux_estimator_results}
\end{figure*}

For a 50-rung flux ladder at $\Phi=0.3\pi$, we keep the amplitude network, sampler, optimizer and training protocol fixed while varying only the phase-gradient estimator. Despite extensive learning-rate optimization and increases in network capacity, the standard estimator remains unable to access the low-error regime. Changing only the estimator instead brings the same compact neural-network representation to subpercent error. The direct and AMVP estimators achieve median errors of $0.50\%$ and $0.43\%$, respectively (Fig.~\ref{fig:flux_estimator_results}a).

The improvement extends beyond the variational energy. Both direct and AMVP recover the DMRG chiral-current profile, with $j_c/j_c^{\mathrm{DMRG}}\approx1.01$, whereas the standard result remains near $0.79$ (Fig.~\ref{fig:flux_estimator_results}b). Estimator choice therefore determines access to the correct phase-sensitive observable, rather than merely lowering the energy.

Notably, increasing neural-network complexity does not remove this barrier: a larger ResNet and wider or deeper MLPs trained with the standard estimator remain trapped in a higher-error regime (Fig.~\ref{fig:flux_estimator_results}a,c). Moreover, the compact AMVP-estimator MLP reaches the low-error state with more than an order of magnitude less GPU time than the standard-estimator ResNet, showing that estimator design can substitute for brute-force increases in model size and computational cost.

To test whether this behavior extends beyond separated amplitude--phase optimization, we further consider an $8\times8$ flux square with periodic boundary conditions using a coupled amplitude--phase neural-network architecture. The shared parameters introduce amplitude--phase correlations into the full stochastic force, which are incorporated by the AMVP variance minimization. In two dimensions, the AMVP estimator enables a compact NQS ansatz to reach energies below the DMRG results obtained up to $\chi=2048$, whereas the standard estimator remains far above this low-error regime
(Fig.~\ref{fig:flux_estimator_results}d). Furthermore, early in training, most AMVP and direct runs reach sub-percent error, whereas no standard run reaches this threshold at any point during optimization (Fig.~\ref{fig:flux_estimator_results}e).

Beyond flux models, we further verify the robustness of the estimator advantage in quantum systems with different origins of complex phase structures (See Supplementary Information Fig.~S1,2), including chiral spin chains\cite{wei2024unveiling} and continuum fractional quantum Hall (FQH) systems\cite{teng2025solving,qian2025describing}. The same trend is observed across these different systems, indicating that the optimization limitation is
associated with stochastic phase-gradient estimation rather than a particular model class.

\textbf{Estimator-limited chemical accuracy.}\enspace Moving beyond flux-driven phases, molecular electronic structure changes the origin of the phase problem. Relative determinant signs arise from correlation-driven interference\cite{choo2020fermionic,barrett2022autoregressive} rather than an externally imposed gauge field; in heavy-element systems, spin--orbit coupling\cite{flad1997spinorbit,melton2016variable} further extends the problem to a broad distribution of complex phases in the working determinant basis.
Because this information is embedded in the many-electron state, the associated gradients are less explicit and more easily obscured by finite-sample fluctuations, yet accurate phase learning remains essential for correlation-energy recovery and chemical accuracy
\cite{barrett2022autoregressive}.

For N$_2$, we perform joint amplitude--phase optimization in a selected-CI determinant space using a shared MLP,
\begin{equation}
\Psi_{\theta}(I)
=
\exp\!\left[
u_{\theta}(I)+\mathrm{i}\varphi_{\theta}(I)
\right],
\qquad I\in\mathcal S.
\end{equation}
For spin-free molecular Hamiltonians, phase optimization primarily learns relative determinant signs. With explicit spin--orbit coupling, the I$_2$ wavefunction instead develops a nontrivial complex phase structure in the determinant basis used by the ansatz: no single global phase renders all coefficients real, and their phases span a broad distribution rather than being restricted to $0$ and $\pi$.
\begin{figure*}[t]
    \centering
    \includegraphics[width=\textwidth]{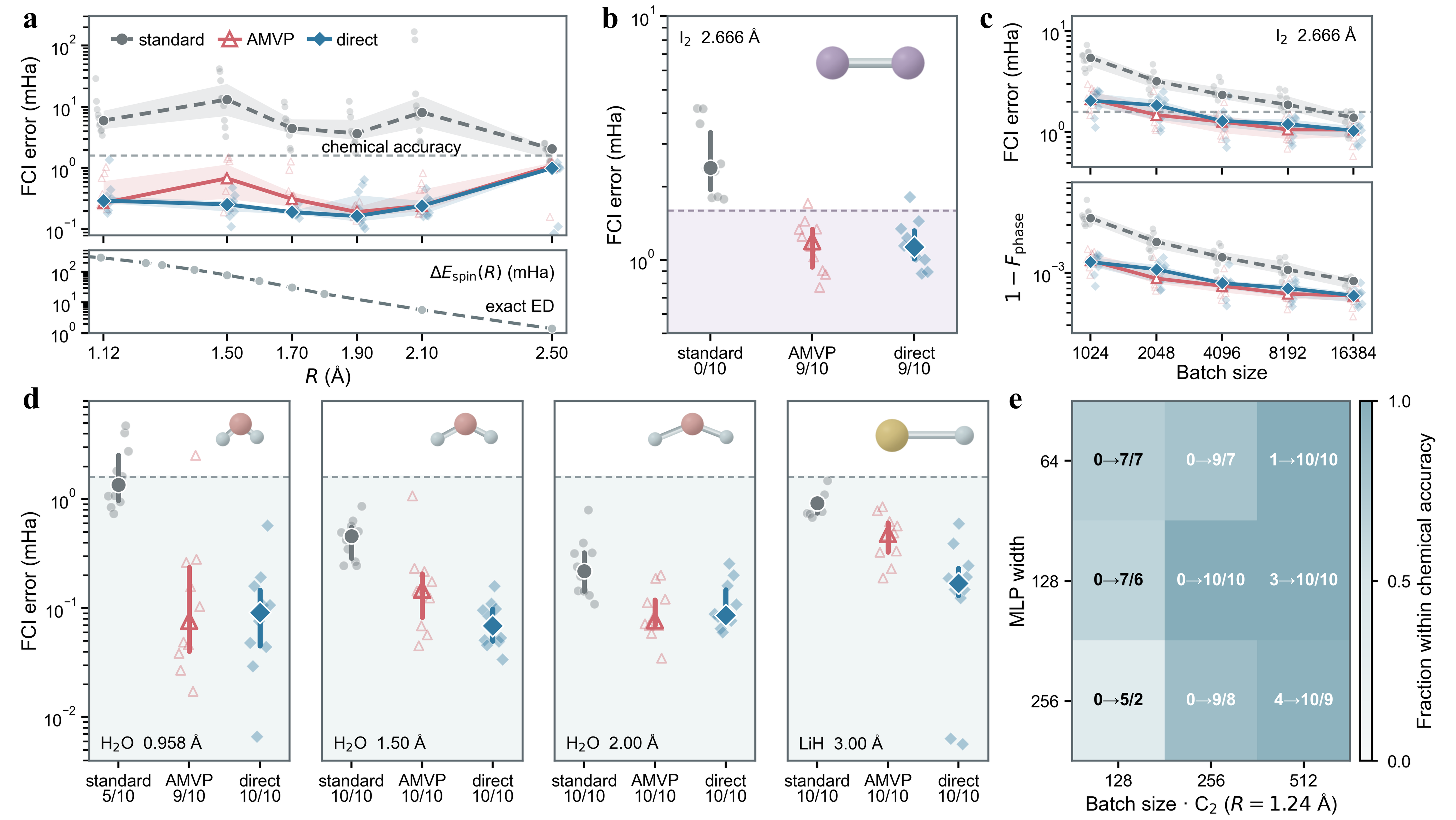}
    \caption{
    \textbf{Estimator-limited trainability of molecular wavefunctions.}
    (a) FCI errors along the N$_2$ bond stretch for the jointly optimized amplitude--phase MLP. The lower strip gives the independently calculated excitation gap from the target-singlet ground state to the first wrong-spin state.
    (b) FCI errors for I$_2$ at $R=2.666~\text{\AA}$ with explicit spin--orbit coupling under joint amplitude--phase optimization.
    (c) Batch-size dependence of fixed-amplitude phase optimization for I$_2$ at $R=2.666~\text{\AA}$ with explicit spin--orbit coupling. The upper and lower panels show the FCI error and phase infidelity
    $1-F_{\mathrm{phase}}$, respectively.
    (d) FCI errors for fixed-amplitude phase-learning controls on H$_2$O at three geometries and LiH.
    (e) Fraction of paired runs within chemical accuracy for C$_2$ at $R=1.24~\text{\AA}$ across MLP widths and batch sizes. Each cell reports success counts in the order standard $\rightarrow$ direct/AMVP; the shading gives the mean direct–AMVP success fraction.
    In (b,d), the left-to-right order is standard, AMVP, direct.
    Small symbols denote individual runs; large symbols show medians; shaded bands and vertical bars indicate interquartile ranges.}
    \label{fig:molecular_phase_benchmarks}
\end{figure*}

Across all six $\mathrm{N}_2$ geometries, direct and AMVP place every paired run within chemical accuracy, whereas the standard estimator fails in every run at the first five geometries and reaches only $3/10$ at the longest bond length (Fig.~\ref{fig:molecular_phase_benchmarks}a). This separation persists as the independently calculated wrong-spin gap collapses from approximately $292$ to $1.4~\mathrm{mHa}$, demonstrating robust access to the target state throughout bond stretching.

The I$_2$ calculation tests the same principle for a heavy-element molecule with explicit spin--orbit coupling (Fig.~\ref{fig:molecular_phase_benchmarks}b,c). With joint amplitude--phase optimization, direct and AMVP reach median full configuration interaction (FCI) errors of $1.13$ and $1.19~\mathrm{mHa}$, respectively, and chemical accuracy in $9/10$ paired runs; standard gives $2.39~\mathrm{mHa}$ and $0/10$ (Fig.~\ref{fig:molecular_phase_benchmarks}b). Under fixed-amplitude phase optimization, at batch size 4096, direct and AMVP reach median FCI errors near $1.3~\mathrm{mHa}$, compared with $2.4~\mathrm{mHa}$ for standard; at the largest batch, both redesigned estimators reach chemical accuracy in all paired runs (Fig.~\ref{fig:molecular_phase_benchmarks}c). The accompanying phase-infidelity reduction shows that the energy gain reflects more accurate complex phase learning.

Furthermore, the $\mathrm{C}_2$ width--batch map separates sampling budget from neural-network capacity (Fig.~\ref{fig:molecular_phase_benchmarks}e). At batch size 256, the standard estimator produces no chemically accurate run at any tested width, whereas direct and AMVP are reliable across the map. Increasing width or batch size alone does not systematically close this gap.

In addition, the same behavior extends across chemically distinct molecular systems (Fig.~\ref{fig:molecular_phase_benchmarks}d). Direct reaches chemical accuracy for all ten runs in every displayed condition and AMVP does so for nine or ten, while standard is less reliable at the equilibrium H$_2$O geometry. Even where standard crosses the threshold, the redesigned estimators further lower the median error. The common pattern across molecules shows that estimators with the same expectation value can differ sharply in the precision and reliability of finite-sample training.

Estimator design therefore reduces the reliance on ever-larger sampling budgets and neural networks in \textit{ab initio} quantum chemistry.

\textbf{Matched-state diagnostics.}\enspace The accuracy improvements in the lattice and molecular calculations raise a mechanistic question: do the estimators possess different stochastic
properties at the same neural-network state, or do their measured statistics merely reflect the fact that training has already carried them to different regions of parameter space? We explain this distinction through matched-state diagnostics.

For an estimator $m$, we denote its total gradient variance by $V_m=\operatorname{tr}[\operatorname{Cov}(\hat g_m)]$. The variance gain and shared-signal batch signal-to-noise ratio (SNR) are
\begin{equation}
\mathcal{R}_{V}^{(m)}
=
\frac{V_{\mathrm{std}}}{V_m},
\qquad
\operatorname{SNR}_{m}
=
\frac{
\sqrt{B}\,\lVert\mu_{\mathrm{common}}\rVert
}{
\sqrt{V_m}
},
\label{eq:variance_and_shared_snr}
\end{equation}
where $B$ is the batch size. The common population signal $\mu_{\mathrm{common}}$ is estimated symmetrically and debiased for finite-batch noise, so its norm is identical for the standard, direct and
AMVP estimators. A variance ratio above unity denotes lower variance than the
standard estimator.

\begin{figure*}[t]
    \centering
    \includegraphics[width=\textwidth]{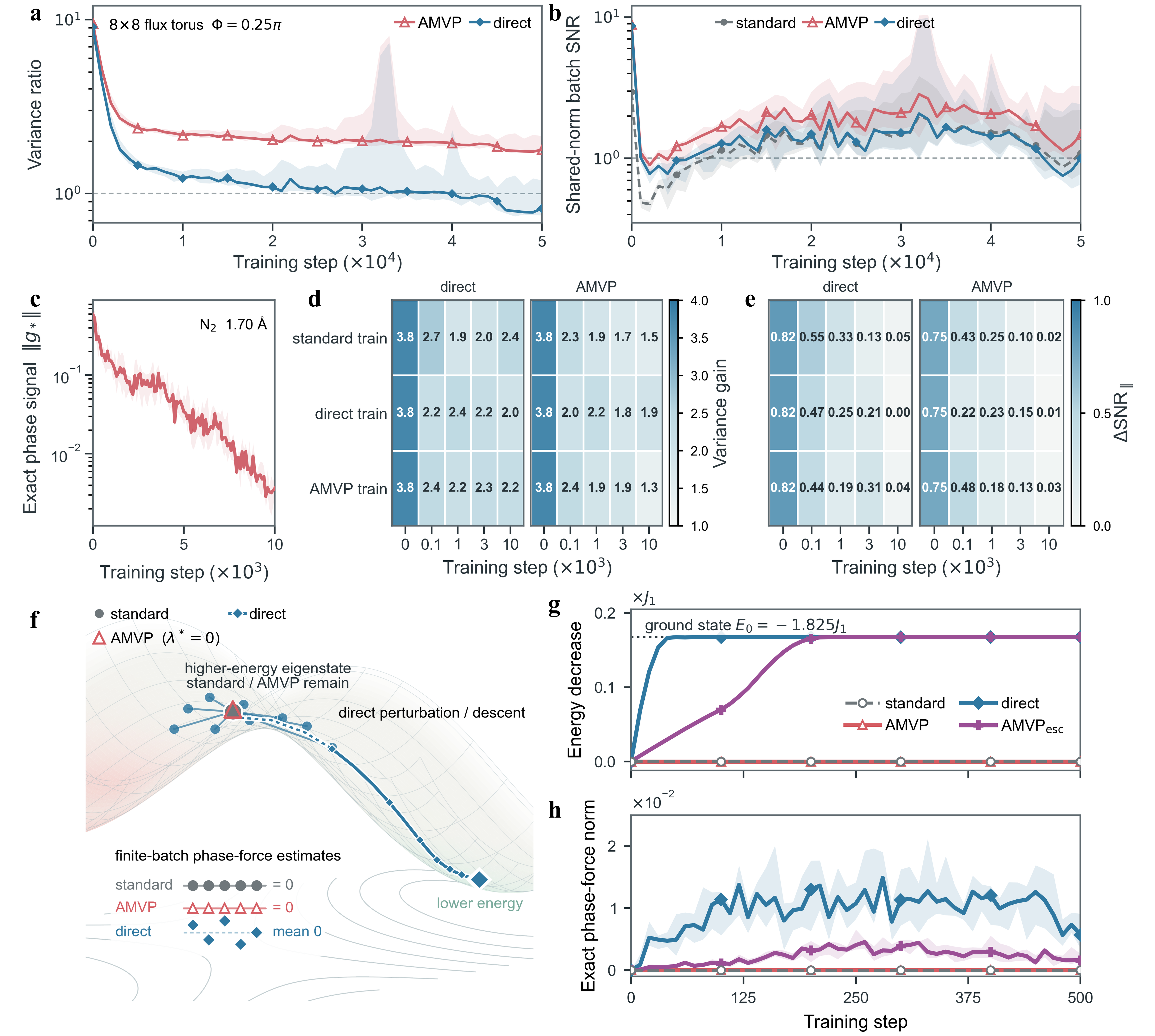}
    \caption{
    \textbf{Matched-state phase-gradient statistics in quantum many-body physics and molecular calculations.}
    (a) Variance ratios $V_{\mathrm{std}}/V_m$
    for the direct and AMVP estimators along the training process of an $8\times8$ flux torus.
    (b) Shared-norm batch SNR for the same flux-torus states, using an identical debiased population-signal estimate for all estimators.
    (c) Exact phase-gradient norm for the N$_2$ fixed-amplitude phase-channel diagnostic at $R=1.70~\text{\AA}$.
    (d) Rows give the estimator that produced the state, columns the checkpoint in thousands of updates; the gains do not depend on which estimator produced the state, showing that the variance difference is a property of the estimator itself rather than of the states.
    (e) Corresponding gains in descent-aligned SNR relative to the standard estimator. 
    (f) Exact-eigenstate control for the periodic honeycomb $J_1$--$J_2$ model. At the higher-energy eigenstate, the standard score-function phase-force estimate vanishes for every finite batch, and AMVP consequently selects $\lambda^*=0$. The direct estimator instead has nonzero finite-batch realizations despite having the same zero population mean, allowing a stochastic displacement toward lower energy. 
    (g,h) Evolution from the same excited eigenstate: energy decrease (g) and exact phase-force norm (h). AMVP$_{\mathrm{esc}}$ denotes the escape-aware extension with a temporary lower bound on the direct mixing weight.
    Curves and bands in (a,b,c,g,h) show medians and interquartile ranges; heatmap cells in (d,e) show medians.
    }
    \label{fig:gradient_diagnostics}
\end{figure*}

At initialization in the $8\times8$ flux-torus calculation, direct and AMVP reduce the gradient variance by roughly ninefold (Fig.~\ref{fig:gradient_diagnostics}a). The relative ordering changes during training and the direct estimator eventually becomes slightly noisier than standard, while AMVP retains the lower-variance channel through its adaptive interpolation.

The shared-signal SNR shows the same trend
(Fig.~\ref{fig:gradient_diagnostics}b): at initialization, it increases from
$2.95$ for the standard estimator to $8.83$ for AMVP, which maintains the highest SNR as the relative estimator variances evolve during training. Because all estimators use the same signal norm, the SNR differences arise entirely from their finite-sample fluctuations.

The phase sector of molecular models is a progressively weaker-signal regime. In the $\mathrm{N}_2$ matched-state fixed-amplitude phase-channel diagnostic, the exact phase-gradient norm decreases from $0.590$ at initialization to $0.00364$ (Fig.~\ref{fig:gradient_diagnostics}c). To measure whether a stochastic gradient remains effective as this signal collapses, we define the descent-aligned SNR
\begin{equation}
\operatorname{SNR}_{\parallel}^{(m)}
=
\frac{
\overline{g}_{m}\cdot g_{\star}
}{
\lVert g_{\star}\rVert\sqrt{V_m}
},
\label{eq:descent_aligned_snr}
\end{equation}
where $g_{\star}$ is the exact full-support phase gradient and $\overline{g}_{m}$ is the minibatch-mean gradient. This quantity rewards both low variance and alignment with the exact descent direction.

Cross-evaluation at fixed checkpoints shows consistently lower direct and AMVP variance across states produced by all three training trajectories (Fig.~\ref{fig:gradient_diagnostics}d). Similar values across the three rows show that the improvement is not restricted to states generated by direct or AMVP training.

The descent-aligned SNR gains are largest during early and intermediate training and decrease as the exact phase force approaches stationarity
(Fig.~\ref{fig:gradient_diagnostics}e). Their seed medians remain non-negative across all three trajectory sources. Evaluating all estimators at the same network checkpoints shows that the
direct and AMVP estimators have lower variance. The gain therefore arises from estimator-specific statistics rather than from differences among the states reached during training.

The connection between these statistics and optimization comes from the standard smoothness bound \cite{bottou2018optimization}. For an unbiased estimator with variance $V$ and an update of size $\eta$ on an $L_{\mathrm{sm}}$-smooth objective,
\begin{equation}
\mathbb{E}[E_{t+1}]-E_t
\leq
-\eta\lVert\nabla E\rVert^2
+
\frac{L_{\mathrm{sm}}\eta^2}{2}
\left(
\lVert\nabla E\rVert^2+V
\right).
\label{eq:variance_descent_bound}
\end{equation}
In the noise-dominated regime
$V\gg\lVert\nabla E\rVert^2$, estimator fluctuations constrain the updates that produce reliable descent. The lower variance therefore raises the descent-aligned SNR at fixed batch size, while leaving the population force and the variational objective unchanged.

Quantum many-body physics and molecular diagnostics reveal the same finite-sample mechanism in two different neural-network state parameterizations. The direct estimator is most effective when weak phase-gradient signals are dominated by standard-estimator noise. AMVP maintains lower variance as the relative ordering changes, adapting between the two unbiased stochastic channels. These diagnostics establish the gradient-quality mechanism underlying the accuracy and reliability improvements in Figs.~\ref{fig:flux_estimator_results} and
\ref{fig:molecular_phase_benchmarks}.

The distinction in Eq.~(\ref{eq:grad_decomposition}) between $\xi_{\mathrm{std}}$ and $\xi_{\mathrm{dir}}$ becomes most transparent at an exact eigenstate, where the population phase force vanishes and only estimator-specific finite-batch fluctuations remain. We demonstrate this in the periodic honeycomb $J_1$--$J_2$ model (Fig.~\ref{fig:gradient_diagnostics}f--h). At the excited Bloch eigenstate, $\operatorname{Im}E_{\mathrm{loc}}=0$ pointwise, so every finite-batch standard phase-force estimate vanishes. The variance-minimizing AMVP weight therefore collapses to the standard endpoint, $\lambda^\star=0$. The direct estimator has the same zero population mean but has nonzero connection-resolved fluctuations on individual batches. A batch realization with a component along a descending direction can perturb the state away from this point. The exact phase force then becomes nonzero and supports continued energy descent. This control reveals a complementary role of estimator fluctuations: variance reduction improves descent in weak-signal regimes, whereas nonzero direct fluctuations can also help optimization leave an excited-state zero-force point.

This analysis motivates an escape-aware AMVP (Fig.~4g,h), in which a temporary lower bound on the direct weight is activated when the standard phase force vanishes but direct finite-batch fluctuations remain. An independent energy evaluation can guide the decision to continue the escape phase or return to variance-minimizing mixing.

\FloatBarrier 

\textbf{Discussion.}\enspace This work establishes gradient-estimator design as a practical route to more accurate and resource-efficient neural-network variational simulations. The origin of phase information changes across the systems studied. Flux phases are externally induced, whereas determinant signs arise from electronic
correlations and spin--orbit coupling produces complex phases in heavy-element molecules. Estimator redesign remains effective throughout this progression, allowing compact neural-network representations to reach low-energy states and chemical accuracy.

Matched-state diagnostics identify the finite-sample origin of these gains. Although the standard and direct estimators share the same population force, estimator redesign lowers gradient variance and raises the signal-to-noise ratio of weak phase directions. The exact-eigenstate analysis reveals a further effect: alternative unbiased estimators can possess different finite-batch zero sets, changing whether stochastic optimization can leave an excited-state zero-force point. Gradient-estimator design therefore shapes both the precision of descent and the stochastic directions available to optimization.

Estimator performance changes during optimization: the direct form is advantageous in weak-signal regimes, whereas the standard estimator becomes zero-variance at exact eigenstates, with AMVP adapting to this crossover.
Their distinct finite-batch zero sets further motivate escape-aware extensions beyond variance minimization alone.

More broadly, these results elevate gradient estimation from a numerical implementation choice to a component of neural-network variational design. The same perspective may apply wherever the same variational gradient admits
alternative finite-sample representations with distinct stochastic properties\cite{kingma2014auto,mohamed2020monte}. Non-Hermitian variational problems provide a further direction, where the Hermiticity identity underlying the standard form no longer holds and finite-sample design can affect both the construction and the efficiency of the variational force. 

\textbf{Methods.}\enspace Architectural and optimization details are provided in Supplementary Information Sec.~1.

\textbf{Variational objective and phase-force estimators.}\enspace For a basis configuration $R$, the Born probability, local energy and variational energy are \cite{foulkes2001qmc,becca2017variational}
\begin{align}
p_\theta(R)&=\frac{|\Psi_\theta(R)|^2}
{\langle\Psi_\theta|\Psi_\theta\rangle},\nonumber\\
E_{\mathrm{loc}}(R)&=\sum_{R'}H_{RR'}
\frac{\Psi_\theta(R')}{\Psi_\theta(R)},\nonumber\\
E(\theta)&=\sum_Rp_\theta(R)\operatorname{Re}E_{\mathrm{loc}}(R).
\end{align}
We write $\Psi_\theta(R)=\exp[u_\theta(R)+\mathrm{i}\varphi_\theta(R)]$
and use $E\equiv E(\theta)$.  For a phase-only parameter $\beta$, one
configuration contributes the amplitude row, the standard phase row and the
direct phase row
\begin{equation}
\begin{aligned}
\hat{\boldsymbol g}_{\mathrm{amp}}(R)
&=2[\operatorname{Re}E_{\mathrm{loc}}(R)-E]\nabla_\theta u_\theta(R),\\
\hat g_\beta^{\mathrm{std}}(R)
&=2\operatorname{Im}E_{\mathrm{loc}}(R)\,\partial_\beta\varphi_\theta(R),\\
\hat g_\beta^{\mathrm{dir}}(R)
&=\operatorname{Re}\partial_\beta E_{\mathrm{loc}}(R)\\
&=\sum_{R'}\operatorname{Im}[W_{RR'}]
[\partial_\beta\varphi_\theta(R)-\partial_\beta\varphi_\theta(R')].
\end{aligned}
\label{eq:phase_force_rows}
\end{equation}
where $W_{RR'}=H_{RR'}\Psi_\theta(R')/\Psi_\theta(R)$ and gradients
through $u_\theta$ are stopped in the direct phase term. Hermiticity gives
\begin{equation}
\langle\hat g_\beta^{\mathrm{dir}}\rangle_{p_\theta}
=2\langle\operatorname{Im}E_{\mathrm{loc}}\,
\partial_\beta\varphi_\theta\rangle_{p_\theta}
=\langle\hat g_\beta^{\mathrm{std}}\rangle_{p_\theta},
\label{eq:direct_standard_identity}
\end{equation}
under the usual differentiability and support conditions.  Thus the raw
estimators have the same population force but different configuration-level
fluctuations.  For a shared real two-head network, the amplitude row in
Eq.~\eqref{eq:phase_force_rows} is retained and only the phase row is
replaced, yielding Eq.~\eqref{eq:coupled_direct_estimator}.

\textbf{Adaptive minimum-variance phase estimation.}\enspace Let $\hat{\boldsymbol g}_{\mathrm{std}}^{\mathrm{ph}}$ and
$\hat{\boldsymbol g}_{\mathrm{dir}}^{\mathrm{ph}}$ denote the two phase
rows.  On the calibration sample we evaluate
$V_s=\operatorname{tr}\operatorname{Cov}
(\hat{\boldsymbol g}_{\mathrm{std}}^{\mathrm{ph}})$,
$V_d=\operatorname{tr}\operatorname{Cov}
(\hat{\boldsymbol g}_{\mathrm{dir}}^{\mathrm{ph}})$, their cross-covariance
trace $C$, and $A_m=\operatorname{tr}\operatorname{Cov}
(\hat{\boldsymbol g}_{\mathrm{amp}},\hat{\boldsymbol g}_{m}^{\mathrm{ph}})$.
Minimizing the trace covariance of the complete force gives
\begin{equation}
\begin{aligned}
\lambda^\star&=\operatorname{clip}_{[0,1]}
\frac{V_s-C+A_s-A_d}{V_s+V_d-2C},\\
\overline{\boldsymbol g}_{\lambda^\star}&=
\overline{\boldsymbol g}_{\mathrm{amp}}+
(1-\lambda^\star)\overline{\boldsymbol g}_{\mathrm{std}}^{\mathrm{ph}}+
\lambda^\star\overline{\boldsymbol g}_{\mathrm{dir}}^{\mathrm{ph}}.
\end{aligned}
\label{eq:amvp_update}
\end{equation}
Here $A_s=A_d=0$ for disjoint amplitude and phase parameter blocks, and overbars are production-batch means.  Hence $\lambda=0$ corresponds to the
standard endpoint and $\lambda=1$ corresponds to the direct endpoint.

\begin{figure}[t]
\refstepcounter{amvpalgorithm}
\label{alg:main}
\rule{\linewidth}{0.8pt}\par
\smallskip
\textbf{Algorithm \theamvpalgorithm.}
\methodname{}: one optimization update.
\par\smallskip
\rule{\linewidth}{0.4pt}
\begin{enumerate}
\setlength{\itemsep}{1.5pt}
\item Draw a production batch for the parameter update. On recalibration steps,
also obtain a calibration batch for estimating $\lambda^\star$.
\item Evaluate $E_{\mathrm{loc}}$, the conventional amplitude force, and the standard and direct phase forces, stopping gradients through $u_\theta$ only when computing the direct phase force.
\item On calibration updates, estimate $V_s,V_d,C$ and, for shared parameters, $A_s,A_d$; evaluate Eq.~\eqref{eq:amvp_update} and retain the
coefficient until the next calibration.
\item Form the complete force using Eq.~\eqref{eq:amvp_update}.  Fixed
endpoints use $\lambda=0$ (standard) or $\lambda=1$ (direct).
\item Apply the common preconditioner or optimizer and update $\theta$.
\end{enumerate}
\vspace{-5pt}
\rule{\linewidth}{0.8pt}
\end{figure}

\textbf{Quantum many-body physics calculations.}\enspace The principal lattice system is the 50-rung flux ladder of
Eq.~\eqref{eq:flux_ladder_ham}.  Independent real networks represent $u_\theta$ and $\varphi_\theta$.  The standard controls vary only the phase-component multiplier after the common preconditioned solve; the MLP and
ResNet capacity controls otherwise use matched estimator protocols.

The ladder reference, $E_{\mathrm{ref}}=-43.302963$, is obtained using finite MPS\cite{white1992density,schollwock2011density} in TeNPy\cite{hauschild2018tenpy} with $\chi=64$ and truncation cutoff $10^{-10}$; doubling $\chi$ changes the energy by $7\times10^{-6}$.  For an oriented bond $i\to j$, the current operator is $\widehat j_{ij}=-\partial H/\partial A_{ij}$, and $j_c(x)=[j_{\mathrm{bottom}}(x)-j_{\mathrm{top}}(x)]/2$.  The scalar current ratio is $\operatorname{median}_{x}
\!\left[
\operatorname{median}_{\mathrm{runs}}j_c(x)/
j_c^{\mathrm{DMRG}}(x)
\right]$ over the displayed bonds. 

The square-flux system contains $8\times8$ sites with periodic boundary condition and $S^z_{\mathrm{tot}}=0$,
\begin{equation}
H_\square=\sum_{\langle ij\rangle}
\left[J_zS_i^zS_j^z+\frac{J_{xy}}{2}
\left(e^{\mathrm{i}A_{ij}}S_i^+S_j^-+\mathrm{h.c.}\right)\right],
\label{eq:square2d_ham}
\end{equation}
with $J_{xy}=1$, $J_z=0$ and flux $\Phi=0.25\pi$ per plaquette.  In Landau gauge, positive vertical bonds carry $A_{(x,y),(x,y+1)}=\Phi x$; horizontal bonds are real except for the periodic-$x$ crossing bond, which carries $-\Phi L_x y$.  From the DMRG energies at $\chi=512,1024,2048$, we set $r=(E_{2048} -E_{1024})/(E_{1024}-E_{512})$ and $E_{\mathrm{ref}}=E_{2048}+(E_{2048}-E_{1024})r/(1-r)\simeq-28.3931$.

The supplementary chiral-chain Hamiltonian is
$H_{\mathrm{ch}}=J\sum_i\mathbf S_i\!\cdot\!\mathbf S_{i+1}
+\alpha\sum_i\mathbf S_i\!\cdot\!(\mathbf S_{i+1}\!\times\!\mathbf S_{i+2})$\cite{wei2024unveiling},
with $J=\alpha=1$, 50 periodic sites and $S^z_{\mathrm{tot}}=0$.
A shared real two-head network is optimized with the same pipeline for all three estimators; the DMRG reference is $-23.748041$. The continuum calculation uses the DeepHall\cite{qian2025describing} complex Psiformer\cite{vonglehn2023selfattention} on the Haldane sphere\cite{haldane1983fractional} at $\nu=1/3$ for $N=6$ ($2Q=15$) and $N=8$ ($2Q=21$).  The corrected energy per particle
is $\frac{E_c}{N}=\sqrt{\frac{2Q\nu}{N}}\,
\frac{E-N/2-N^2/(2\sqrt Q)}{N}$ in units of $e^2/(\epsilon\ell_B)$.  

\textbf{Molecular phase optimization.}\enspace For N$_2$ in Fig.~\ref{fig:molecular_phase_benchmarks}a and Supplementary Information Fig.~S3a,c, amplitude and phase are optimized jointly on a fixed determinant support. For I$_2$, Fig.~\ref{fig:molecular_phase_benchmarks}c isolates phase optimization at fixed amplitude, whereas Fig.~\ref{fig:molecular_phase_benchmarks}b optimizes amplitude and phase jointly under the same explicitly spin--orbit-coupled Hamiltonian.

The spin gap\cite{li2024spinsymmetry} in Fig.~\ref{fig:molecular_phase_benchmarks}a is the energy difference between the target-singlet ground state and the lowest excited state with a different total spin.

\textbf{Matched-state diagnostics and exact-state control.}\enspace The square-torus diagnostics evaluate all phase-force estimators on the same $B$ configurations using the variance and shared-signal SNR definitions
in Eq.~\eqref{eq:variance_and_shared_snr}.

The exact-eigenstate analysis in Fig.~\ref{fig:gradient_diagnostics}f--h uses the periodic one-particle honeycomb $J_1$--$J_2$ model,
\begin{equation}
\hat H
=
-J_1\sum_{\langle i,j\rangle}
\left(\hat c_i^\dagger \hat c_j+\mathrm{h.c.}\right)
+J_2\sum_{\langle\!\langle i,j\rangle\!\rangle}
\left(\hat c_i^\dagger \hat c_j+\mathrm{h.c.}\right).
\label{eq:honeycomb_j1j2}
\end{equation}


\textbf{Statistical analysis.}\enspace All comparisons use ten paired independent runs and are summarized by the median and interquartile range.

\textbf{Data availability.}\enspace The data supporting the findings of this study are publicly available in the AMVP repository\cite{chenanphys2026amvp}.

\textbf{Code availability.}\enspace The associated source code and calculation workflows are publicly available in the AMVP repository\cite{chenanphys2026amvp}.

\textbf{Acknowledgements.}\enspace We acknowledge Yubing Qian, Xiaoyong Ni, Xin Jin, Ji-Zhong Jiang and Khachatur Nazaryan for fruitful discussions. We also thank Ji Chen's group for making the DeepHall code available, parts of which served as a reference for our implementation. This work was supported by the National R\&D Program of China (2024YFA1410500, 2022YFA1403601), the Innovation Program for Quantum Science and Technology (Grant No. 2021ZD0302800), the National Natural Science Foundation of China (No. 12322402, No. 12274206), the Natural Science Foundation of Jiangsu Province (No. BK20233001), the Fundamental Research Funds for the Central Universities (No. KG202501), the Armenian Higher Education and Science Committee ARPI Remote Laboratory program 24RL-1C024, research projects 21AG-1C024 and 25Post-Doc1C003.

\textbf{Competing interests.}\enspace The authors declare no competing interests.

\renewcommand{\bibfont}{\fontsize{9pt}{10pt}\selectfont}
%

\end{document}